\documentstyle[twocolumn,aps]{revtex}

\tightenlines

\newcommand{\beq}{\begin{equation}}
\newcommand{\eeq}{\end{equation}}
\newcommand{\beqa}{\begin{eqnarray}}
\newcommand{\eeqa}{\end{eqnarray}}
\newcommand{\ahat}{\hat{a}}
\newcommand{\bhat}{\hat{b}}
\newcommand{\adag}{\hat{a}^\dagger}
\newcommand{\bdag}{\hat{b}^\dagger}
\newcommand{\den}{\hat{\rho}}
\newcommand{\Lhat}{\hat{L}}
\newcommand{\Nhat}{\hat{N}}
\newcommand{\Uhat}{\hat{U}}

\newcommand{\com}[1]{\left[ #1 \right]}
\newcommand{\prt}[1]{\left( #1 \right)}
\newcommand{\calH}{{\cal H}}
\newcommand{\ket}[1]{| #1 \rangle}
\newcommand{\bra}[1]{\langle #1 |}
\newcommand{\qqquad}{\quad \quad \quad}
\newcommand{\refeq}[1]{$\prt{\ref{eq:#1}}$}
\newcommand{\identity}{\hat{\openone}}

\begin{document}

\draft

\title{Unpolarized light in quantum optics}

\author{Jonas S\"{o}derholm,$^{1,*}$ Gunnar Bj\"{o}rk,$^1$ and Alexei Trifonov$^{1,2}$}

\address {$^1$Department of Electronics, Royal Institute of Technology (KTH), Electrum 229, SE-164 40 Kista, Sweden\\
$^2$Ioffe Physical Technical Institute, 26 Polytekhnicheskaya,
194021 St. Petersburg, Russia\\
$^*$Electronic address: jonas@ele.kth.se}

%\date{\today}

\maketitle

\begin{abstract}
We present a new derivation of the unpolarized quantum states of light, whose general form was first derived by Prakash and Chandra [Phys. Rev. A {\bf 4}, 796 (1971)]. Our derivation makes use of some basic group theory, is straightforward, and offers some new insights.
\end{abstract}


%%%%%%%%%%%%%%%%%%%%%%%%%%%%%%%%%%%%%%%%%%%%%%%%%%%%%%%%%%%%%%%%%%%%%%%%%%%%

\section{Introduction}

The quantum theory of polarization formally has the same structure as that for relative phase. This follows from the fact that both observables measure the relative properties of two bosonic modes. As is well known, the phase shift is generated by one of Schwinger's bosonic SU(2) operators \cite{Schwinger}, which form the Lie algebra for linear lossless transformations of two bosonic modes \cite{Campos}. Similarly, when considering polarization another SU(2) operator appears, generating geometric rotations. This makes it possible to treat the polarization in the same manner as was successfully applied to the relative phase \cite{Luis,Bjork,Trifonov}, that is, considering the different excitation manifolds separately. In this way we are able to easily find all the unpolarized quantum states of light first derived by Prakash and Chandra \cite{Prakash}, and later treated by Agarwal \cite{Agarwal} and by Lehner, Leonhardt, and Paul \cite{Lehner}. We think our treatment offers some insight into the nature of polarization at a quantum level.

\section{Schwinger's bosonic SU(2) operators}

We will consider a transverse field of light (or more generally, a transverse field of bosons) that can be decomposed into two linearly polarized modes, called $a$ and $b$. The field is supposed to have a well-defined direction of propagation. Assuming that the two modes have the same frequency, any lossless coupling will leave the number of excitations (photons) unchanged. It is therefore natural to introduce Schwinger's realization of the SU(2) algebra \cite{Schwinger}
\beqa \Lhat_1 & = & \frac{1}{2}(\adag \bhat + \ahat \bdag) , \\
\Lhat_2 & = & \frac{1}{2 i}(\adag \bhat - \ahat \bdag) , \\
\Lhat_3 & = & \frac{1}{2}(\adag \ahat - \bdag \bhat) , \eeqa
where $\ahat$ and $\adag$ denote the annihilation and creation operator of mode $a$, respectively. The SU(2) operators satisfy
\beq \com{\Lhat_k,\Lhat_l} = i \epsilon_{klm} \Lhat_m , \label{eq:Liealg} \eeq
where the Levi-Civit\`{a} tensor $\epsilon_{klm}$ is equal to $1$ and $-1$ for even and odd permutations of its indices, respectively, and zero otherwise. The Casimir operator is
\beq \Lhat^2 \equiv \Lhat_1^2 + \Lhat_2^2 + \Lhat_3^2 = \frac{\Nhat}{2} \prt{\frac{\Nhat}{2}+1} , \eeq
where $\Nhat = \adag \ahat + \bdag \bhat$ is the total photon number. Thus $\Nhat$ commutes with all the generators $\Lhat_k, k=1,2,3$, which shows that the coupling described by them are indeed lossless. The Schwinger operators $\Lhat_1$, $\Lhat_2$, $\Lhat_3$, and $\Lhat^2$ are proportional to the Stokes operators $\hat{S}_x$, $\hat{S}_y$, $\hat{S}_z$, and $\hat{S}^2$, that can be constructed by quantizing the Stokes parameters used to characterise the polarisation state of classical light fields. 

The evolutions under the SU(2) algebra span all the linear lossless transformations of two bosonic modes \cite{Campos} and they can be written as
\beq \Uhat \prt{\phi_1,\phi_2,\phi_3} = e^{i (\phi_1 \Lhat_1 + \phi_2 \Lhat_2 + \phi_3 \Lhat_3)} , \label{eq:Uhat} \eeq
where $\phi_1, \phi_2$, and $\phi_3$ are real. In physical terms, any combination of differential phase shifts and rotations around the direction of propagation can be expressed in this form using the appropriate variables $\phi_k$. Since all the generators commute with the total photon number, any evolution takes the form of a direct sum
\beq \Uhat \prt{\phi_1,\phi_2,\phi_3} = \mathop{\oplus}_{N=1}^\infty \Uhat_N \prt{\phi_1,\phi_2,\phi_3} , \eeq
where $\Uhat_N$ operates on the $N$-dimensional Hilbert space $\calH_N$ formed by all the two-mode states with $N-1$ photons. This means that the matrix corresponding to $\Uhat \prt{\phi_1,\phi_2,\phi_3}$ is block-diagonal if the basis is chosen to be the two-mode number states ordered by their excitation manifold, for example, ($\ket{0,0}$, $\ket{0,1}$, $\ket{1,0}$, $\ket{0,2}$, $\ket{1,1}$, $\ket{2,0}$, $\ket{0,3}$, $\ldots$). Similarly, the Hilbert space of two harmonic oscillators can be written
\beq \calH_\infty \otimes \calH_\infty = \mathop{\oplus}_{N=1}^\infty \calH_N . \eeq

\section{Unpolarized quantum states}

Prakash and Chandra argued that the unpolarized quantum states should be defined as those states that are invariant under any geometric rotation and any phase shift. From these conditions the most general form of an unpolarized quantum state was derived \cite{Prakash}. An equivalent definition, but using different derivations, lead Agarwal, and later Lehner {\it et al.}, to arrive at the same result \cite{Agarwal,Lehner}. Here we present a fourth, alternative, more straightforward, derivation, which makes use of Shur's lemma.

In mathematical form the invariance requirements can be written
\beq \com{\den,e^{i \theta_k \Lhat_k}} = 0 , \qqquad \forall \, \theta_k , \qqquad k = 2, 3, \eeq
where $\den$ is the density operator of the quantum state. Since this is valid for all $\theta_2$ and $\theta_3$, we have
\beq \com{\den,\Lhat_k} = 0 , \qqquad k = 2 ,3 . \label{eq:cond} \eeq

We start by observing that $\Lhat_1$ can be expressed in $\Lhat_2$ and $\Lhat_3$, as is seen from Eq. \refeq{Liealg}. Thus, any operator that commutes with both $\Lhat_2$ and $\Lhat_3$ also commutes with $\Lhat_1$. Therefore the unpolarized states satisfy
\beq \com{\den,\Lhat_k} = 0 , \qqquad k = 1, 2 ,3 . \eeq
This implies, according to Shur's lemma \cite{Shur}, that the density operator is a multiple of the identity operator in the respective Hilbert spaces $\calH_N$. Since density operators are Hermitian, we have
\beq \den = \mathop{\oplus}_{N=1}^\infty r_N \identity_N , \label{eq:unpolst} \eeq
where $\identity_N$ is the identity operator in $\calH_N$, and all $r_N$ are real and positive. Writing this equation explicitly gives
\beq
\den = r_1 \ket{0,0}\bra{0,0}+r_2 (\ket{0,1}\bra{0,1}+\ket{1,0}\bra{1,0}) + \ldots 
\eeq
Furthermore, the trace of any density operator equals unity, so we find
\beq \sum_{N=1}^\infty N \, r_N = 1 . \eeq
We note that {\em the vacuum state is the only pure state that is unpolarized}, and that {\em the unpolarized mixed states are totally mixed in each manifold}. Any two-mode thermal state is hence unpolarized.

It is clear that Eq. \refeq{cond} is equivalent to
\beq \com{\den,\Uhat \prt{\phi_1,\phi_2,\phi_3}} = 0 , \qqquad \forall \, \phi_1, \phi_2, \phi_3 , \eeq
where $\Uhat \prt{\phi_1,\phi_2,\phi_3}$ denotes the linear lossless transformations \refeq{Uhat}. Moreover, the form of the unpolarized states \refeq{unpolst} makes them invariant under any, not necessarily linear, lossless transformation.

\section{Discussion}

We have investigated the quantum mechanical definition of unpolarized light \refeq{cond}, suggested by Prakash and Chandra \cite{Prakash}. (The choice $k=1,2$, or $k=1,3$, leads to equivalent definitions.) The suggested definition is very concise and can experimentally be stated as: An unpolarized state is invariant under any rotation around its direction of propagation and any differential phase shift, or any combination thereof. Note that this definition is more severe than the classical definition of unpolarized light, as pointed out by Agarwal \cite{Agarwal}. Classically, the term ``unpolarized'' only implies invariance of the second order moments of the two modes under differential phase shifts and geometrical rotations. The quantum mechanical definition implies invariance of the moments of all orders. 

It is somewhat surprising that such a relatively ``mild'' defining condition \refeq{cond} leads to such severe restraint on the density matrix as Eq. (\ref{eq:unpolst}). If one defines the circularly polarized states as the states that are invariant under rotation around the direction of propagation, one finds that there are $N+1$ such (orthonormal) states in every excitation manifold $N$. These states can be used to form a Hermitian circular polarization operator in a similar manner as the relative phase operator has been constructed \cite{Luis,Bjork}. In the $N$th excitation manifold one can subsequently construct up to $N$ additional, complementary, operators \cite{Wootters}. An alternative definition of an unpolarized transverse state would be that such a state should be invariant under {\em all} transformations generated by these complementary operators. A direct check of what requirement on the density matrix this alternative (but in our eyes seemingly stronger) definition of unpolarized light dictates, again leads to Eq. (\ref{eq:unpolst}). Since Eq. (\ref{eq:cond}) is commendably concise, and is experimentally testable using only linear components (only variable phase-plates are needed), we think it is a correct and convenient mathematical definition of unpolarized light.

\section{Acknowledgments}
This work was funded by the Swedish Foundation for Strategic
Research (SSF), the Royal Swedish Academy of Science (KVA), and the Swedish Research Council for Engineering Sciences (TFR).

%%%%%%%%%%%%%%%%%%%%%%%%%%%%%%%%%%%%%%%%%%%%%%%%%%%%%%%%%%%%%%%%%%%%%%%%%%%%

\end{document}